\documentclass{article}

\usepackage{arxiv}

\usepackage[utf8]{inputenc} 
\usepackage[T1]{fontenc}    
\usepackage[hidelinks]{hyperref}       
\usepackage{url}            
\usepackage{booktabs}       
\usepackage{amsfonts, amsmath, amssymb}   
\usepackage{nicefrac}       
\usepackage{microtype}      
\usepackage{lipsum}		
\usepackage{graphicx}
\usepackage{mathtools}
\usepackage{multirow}
\usepackage{tabularx}
\usepackage{rotating}
\usepackage{natbib}
\usepackage{doi}
\usepackage{xcolor}

\def\nSDR{$\mathrm{nSDR}$}

\title{Toward Deep Drum Source Separation}

\author{\href{https://orcid.org/0000-0002-9438-3190}{\includegraphics[scale=0.065]{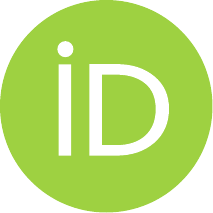}\hspace{1mm}Alessandro Ilic Mezza},\quad \href{https://orcid.org/0000-0001-6144-8288}{\includegraphics[scale=0.065]{orcid.pdf}\hspace{1mm}Riccardo Giampiccolo},\quad \href{https://orcid.org/0000-0001-7973-0134}{\includegraphics[scale=0.065]{orcid.pdf}\hspace{1mm}Alberto Bernardini},\quad \href{https://orcid.org/0000-0002-5803-1702}{\includegraphics[scale=0.065]{orcid.pdf}\hspace{1mm}Augusto Sarti}\vspace{1mm}\\
        Dipartimento di Elettronica, Informazione e Bioingegneria\\
	Politecnico di Milano\\
	Milan, Italy \\
	\texttt{alessandroilic.mezza@polimi.it},\; \texttt{riccardo.giampiccolo@polimi.it},\\ 
        \texttt{alberto.bernardini@polimi.it},\; \texttt{augusto.sarti@polimi.it}\\
}

\date{}

\renewcommand{\shorttitle}{Toward Deep Drum Source Separation}

\hypersetup{
pdftitle={Toward Deep Drum Source Separation, Mezza et al., 2023},
pdfsubject={eess.AS, cs.LG, cs.SD},
pdfauthor={Alessandro Ilic Mezza, Riccardo Giampiccolo, Alberto Bernardini, Augusto Sarti},
pdfkeywords={Deep learning, Drums, Source Separation},
}

\begin{document}
\maketitle

\begin{abstract}
In the past, the field of drum source separation faced significant challenges due to limited data availability, hindering the adoption of cutting-edge deep learning methods that have found success in other related audio applications. In this manuscript, we introduce StemGMD, a large-scale audio dataset of isolated single-instrument drum stems. Each audio clip is synthesized from MIDI recordings of expressive drums performances using ten real-sounding acoustic drum kits. Totaling 1224 hours, StemGMD is the largest audio dataset of drums to date and the first to comprise isolated audio clips for every instrument in a canonical nine-piece drum kit. We leverage StemGMD to develop LarsNet, a novel deep drum source separation model. Through a bank of dedicated U-Nets, LarsNet can separate five stems from a stereo drum mixture faster than real-time and is shown to significantly outperform state-of-the-art nonnegative spectro-temporal factorization methods.
\end{abstract}


\vspace{2em}
\centerline{\Large\bf \color{red} Peer-Reviewed Version}

{\color{red} A peer-reviewed version of the manuscript has been published in \textit{Pattern Recognition Letters} and is available at \href{http://dx.doi.org/10.1016/j.patrec.2024.04.026}{10.1016/j.patrec.2024.04.026}. \bf Please, cite this work as: A. I. Mezza, R. Giampiccolo, A. Bernardini, and A. Sarti, "Toward Deep Drum Source Separation," \textit{Pattern Recognition Letters}, vol. 183, pp. 86-91, 2024, doi: 10.1016/j.patrec.2024.04.026.\\

\texttt{@article\{mezza2024,\\
title = \{Toward deep drum source separation\},\\
author = \{Alessandro Ilic Mezza and Riccardo Giampiccolo and Alberto Bernardini and Augusto Sarti\},\\
journal = \{Pattern Recognition Letters\},\\
volume = \{183\},\\
pages = \{86-91\},\\
year = \{2024\},\\
issn = \{0167-8655\},\\
doi = \{https://doi.org/10.1016/j.patrec.2024.04.026\}\\
\}}
}%

\section{Introduction}
\label{sec:introduction}
Deep learning has been a breakthrough in the field of acoustic source separation, with applications ranging from speech~\citep{speech2018overview} to a variety of environmental sounds~\citep{kavalerov2019universal}.
In recent years, neural-network-aided music demixing models (MDX) have shown tremendous results. Open-Unmix~\citep{stoter19}, Spleeter~\citep{hennequin2020spleeter}, Meta-TasNet~\citep{meta-tasnet2020}, CrossNet-UMX~\citep{sawata2021all}, and Demucs~\citep{defossez2021hybrid}, to name a few, have indeed proved very capable at extracting various stems from a mixed music track.
MDX models are typically trained to isolate four stems~\citep{SiSEC2018, mdx2021, mdx2023}, i.e., ``vocals'', ``bass'', ``drums'', and ``other'', while sometimes including also other instruments such as guitar and piano. 
In practice, however, the drum kit is conventionally treated as a single instrument despite being an ensemble of various percussive instruments in itself. Unfortunately, this prevents the full inversion of the mixing process performed by the sound engineer when producing a song. 
High-quality Drum Source Separation (DSS) could have a significant impact on many creative applications, as it may enhance remixing, remastering, and audio production by providing precise control over individual drum
element, allowing artists and sound engineers to fine-tune their tracks beyond the capabilities of existing MDX
software. Additionally, it can be valuable in music
analysis, facilitating detailed studies of drum patterns and
rhythms, as well as in music education to isolate drum parts
for instructional purposes. Furthermore, DSS may be instrumental in advancing Automatic Drum Transcription (ADT)~\citep{wu2018review}.

Deep learning methods are data-hungry. However, as shown in Table \ref{tab:datasets}, available datasets of isolated drum stems are few and of limited size.
This substantial lack of data is possibly one of the main reasons why, so far, deep DSS has not been investigated in the literature, unlike other drum-related tasks such as, e.g., ADT for which deep learning approaches have been studied extensively~\citep{vogl2017drum, choi2019deep, eGMD, ishizuka2021global}.
By contrast, the latest advances in DSS to date are based on variants of nonnegative matrix factorization (NMF) and nonnegative matrix factor deconvolution (NMFD), as they prove effective even when only a small amount of data is available~\citep{dittmar2014real, dittmar2016amen, vande2021sigmoidal, cai2021dual}. 

With this work, our goal is to address the longstanding data scarcity problem that held back research into more advanced data-driven DSS methods in the past few decades, and demonstrate that deep neural networks are not only a viable solution but also an efficient tool when it comes to DDS.
In this manuscript, we present StemGMD, a new large-scale audio dataset of isolated drum stems synthesized from several hours of expressive performances.
We leverage StemGMD to develop LarsNet, a deep DSS model that separates five stems from a stereo drum mixture. Each stem is extracted by a dedicated U-Net yielding a spectro-temporal soft mask that is applied to the short-time Fourier transform (STFT) of the mixture signal.
In our experiments, LarsNet consistently outperform state-of-the-art DSS methods based on NMF and NMFD, while greatly reducing inter-channel cross-talk artifacts\footnote{Audio examples are available at \url{https://polimi-ispl.github.io/larsnet}} and computational time.

\begin{table}[t]
    \centering
    \caption{Overview of existing drums datasets.}
    \resizebox{\linewidth}{!}{%
    \begin{tabular}{lcccccccccc}
    \toprule
         & \textbf{Clips} & \textbf{Duration (h)} & \textbf{Classes/Mics} & \textbf{Drum kits} & \textbf{Real/Synth} & \textbf{Human} & \textbf{Mixture} & \textbf{Transcription} & \textbf{Isolated stems} \\
         \midrule
         MDB-Drums~\citep{southall2017mdb} & $23$ & $0.345$ & $21$ & $\le 23$ & R & $\checkmark$ & Mono & TXT & $\times$ \\
         IDMT-SMT-Drums~\citep{dittmar2014real} & $608$ & $2.1$ & $3$ & N/A & R/S & $\times$ & Mono &  XML & $64$ mixtures\\
         ENST-Drums~\citep{gillet2006enst} & $456$ & $3.75$ & $20/8$ & $3$ & R & $\checkmark$ & Stereo & TXT & $\checkmark$ \\
         GMD~\citep{groove2019} & $1150$ & $13.6$ & $22$ & $1$ & S & $\checkmark$ & Mono & MIDI & $\times$ \\
         TMIDT~\citep{vogl2018towards} & $4197$ & $259$ & $18$ & $57$ & S & $\times$ & Mono &  TXT & $\times$ \\
         E-GMD~\citep{eGMD} & $45\,537$ & $444.5$ & $22$ & $43$ & S & $\checkmark$ & Mono & MIDI & $\times$ \\
         \midrule
        StemGMD (ours) & $103\,500$ & $1224$ & $9$ & $10$ & S & $\checkmark$ & Stereo & MIDI & $\checkmark$\\
        \bottomrule
    \end{tabular}%
    }%
    \label{tab:datasets}
\end{table}

\section{StemGMD}
\label{sec:dataset}
In 2019, Magenta released GMD, a large corpus comprising 13.6 hours of human-performed drum tracks recorded by ten drummers playing on a Roland TD-11 electronic drum kit~\citep{groove2019}. GMD contains MIDI files for each performance, along with the corresponding full-kit audio mixtures. Later on, \citet{eGMD}~introduced E-GMD, an ADT dataset containing about 444~hours of audio data gathered by re-recoding all GMD sequences using 43 drum kits.
Despite being valuable for many tasks, such as beat generation, drum infilling, ADT, and groove transfer \citep{groove2019}, these datasets do not contain isolated stems and are thus unsuited for training deep source separation models. In this work, we expand the GMD family by introducing StemGMD, a new large-scale corpus of isolated drum stems.

First, we applied the note mapping proposed in~\citep{groove2019} to the raw MIDI data provided with GMD.
Such mapping, summarized in Table \ref{tab:groove_mapping},
preserves the velocity and timing of each note while reducing the 22 original MIDI channels to nine canonical instruments (whose General MIDI note numbers are reported in brackets), i.e.,
Kick Drum~(36),
Snare~(38),
High Tom~(50),
Low-Mid Tom~(47),
High Floor Tom~(43),
Closed Hi-Hat~(42),
Open Hi-Hat~(46),
Crash Cymbal~(49),
and
Ride Cymbal~(51).

Then, we synthesized each of the nine channels of the resulting MIDI clips as 16-bit/$44.1$~kHz stereo WAV files. We used high-fidelity Logic Pro X drum libraries, selecting ten different acoustic drum kits in order to cover a wide range of timbres. 
To ensure the superposition principle that underlies most source separation techniques, we obtained the mixture signals simply by summing the nine synthesized stems. 
This procedure preserves a one-to-one correspondence between the audio tracks and the original MIDI files. Hence, GMD metadata were kept for each clip, e.g., duration, genre, tempo, and drummer ID, which we augmented with the drum kit ID. The resulting audio dataset, which we call StemGMD, is made freely available online.\footnote{StemGMD: \url{https://zenodo.org/records/7860223}} 

StemGMD comprises more than 136~hours of drum mixtures and totals approximately 1224~hours of audio data. 
With 103\,500 clips, to the best of our knowledge, StemGMD is the largest publicly available dataset of drums. Moreover, it is the first large-size dataset of isolated stems to account for all pieces in a standard drum kit, including toms and cymbals. 

The train, test, and validation folds from Magenta's GMD are retained in StemGMD. However, the recommended test fold amounts to more than ten hours. Therefore, in this work, we use a more manageable test split named \textit{Eval Session}. StemGMD \textit{Eval Session} comprises 400 drum mixtures (40 for each drum kit) and spans several genres, as it was collected by tasking four drummers to record the same set of ten beats in their own style~\citep{groove2019}.

\begin{table}[t]
    \centering
    \caption{MIDI note mapping \citep{groove2019}.}
    \label{tab:groove_mapping}
    \resizebox{0.6\columnwidth}{!}{%
    \begin{tabular}{ccccc} 
    \toprule 
     \textbf{Original Pitch}&  \textbf{Roland Mapping}
&\textbf{General MIDI Mapping} &  \textbf{StemGMD Mapping} &\textbf{StemGMD Pitch}
\\ \midrule 
     36&   Kick
&Bass Drum 1 &  Kick Drum &36
\\ \midrule 
     38&   Snare (Head)
&Acoustic Snare &  Snare &38
\\ \midrule 
     40&   Snare (Rim)
&Electric Snare &  Snare &38
\\ \midrule 
     37&   Snare X-Stick
&Side Stick &  Snare &38
\\ \midrule 
     48&   Tom 1
&Hi-Mid Tom &  High Tom &50
\\ \midrule 
     50&   Tom 1 (Rim)
&High Tom &  High Tom &50
\\ \midrule 
     45&   Tom 2
&Low Tom &  Low-Mid Tom &47
\\ \midrule 
     47&   Tom 2 (Rim)
&Low-Mid Tom &  Low-Mid Tom &47
\\ \midrule 
     43&   Tom 3 (Head)
&High Floor Tom &  High Floor Tom &43
\\ \midrule 
     58&  Tom 3 (Rim)
&Vibraslap & High Floor Tom &43
\\ \midrule 
     46&  HH Open (Bow)
&Open Hi-Hat & Open Hi-Hat &46
\\ \midrule 
     26&  HH Open (Edge)
&N/A & Open Hi-Hat &46
\\ \midrule 
     42&  HH Closed (Bow)
&Closed Hi-Hat & Closed Hi-Hat &42
\\ \midrule 
     22&  HH Closed (Edge)
&N/A & Closed Hi-Hat &42
\\ \midrule 
     44&  HH Pedal
&Pedal Hi-Hat & Closed Hi-Hat &42
\\ \midrule 
     49&  Crash 1 (Bow)
&Crash Cymbal 1 & Crash Cymbal &49
\\ \midrule 
     55&  Crash 1 (Edge)
&Splash Cymbal & Crash Cymbal &49
\\ \midrule 
     57&  Crash 2 (Bow)
&Crash Cymbal 2 & Crash Cymbal &49\\ \midrule 
     52&  Crash 2 (Edge)
&Chinese Cymbal & Crash Cymbal &49\\ \midrule 
     51&  Ride (Bow)
&Ride Cymbal 1 & Ride Cymbal &51\\ \midrule 
     59&  Ride (Edge)
&Ride Cymbal 2 & Ride Cymbal &51\\ \midrule 
     53&  Ride (Bell)
&Ride Bell & Ride Cymbal &51\\ 
\bottomrule
    \end{tabular}
    }
\end{table}

\section{LarsNet}
This section presents a new deep DSS model dubbed LarsNet,\footnote{Pretrained LarsNet models are available at \url{https://github.com/polimi-ispl/larsnet}} which is designed to separate five drum stems from a stereo drum mixture, i.e., kick drum (KD), snare drum (SD), tom-toms (TT), hi-hat (HH), and cymbals (CY). More specifically, TT includes High Tom, Low-Mid Tom, and High Floor Tom; HH includes Open and Closed Hi-Hat; and CY includes Crash and Ride cymbals. 

\subsection{Model Architecture}
\begin{figure}[t]
 \centering 
 \includegraphics[width=0.53\linewidth]{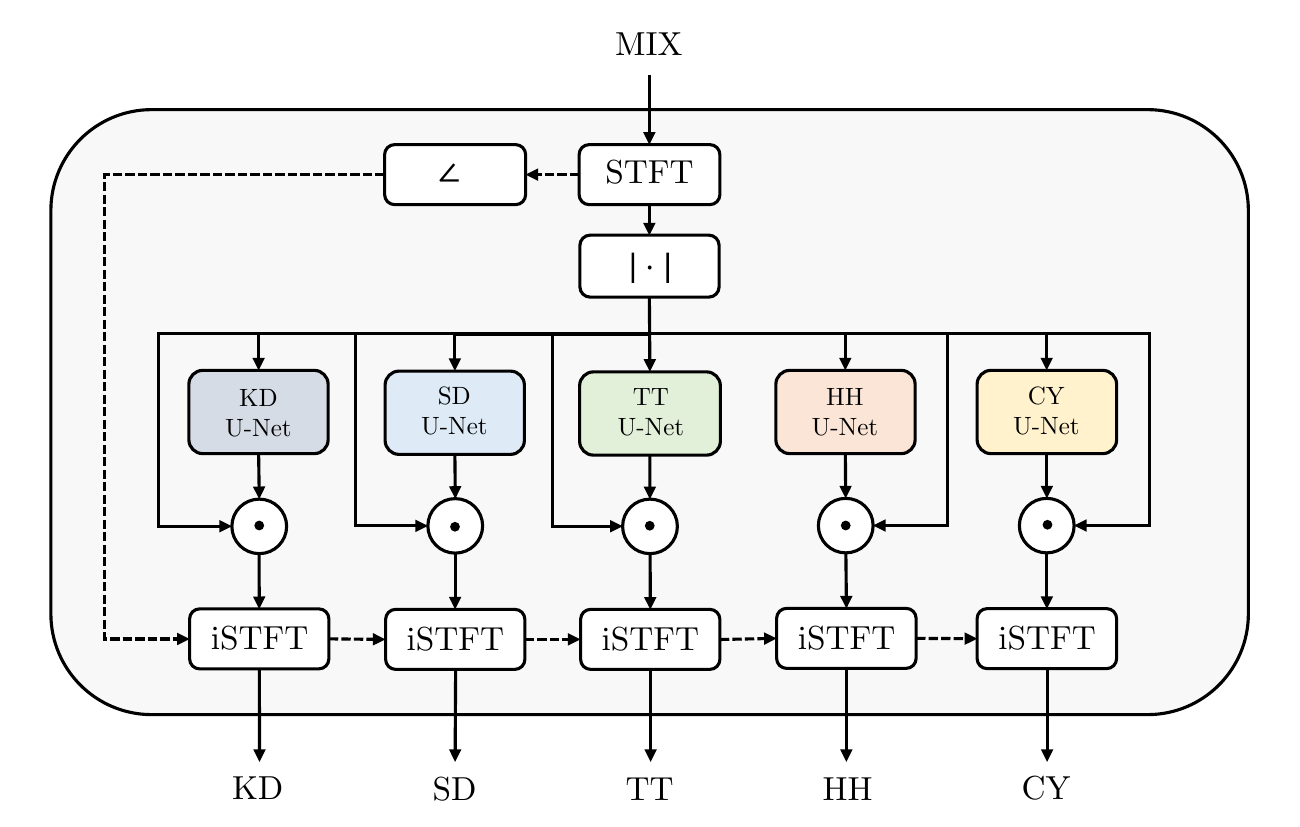}
 \caption{LarsNet architecture.}
 \label{fig:larsnet}
\end{figure}

\begin{figure}[t]
 \centering 
 \includegraphics[width=0.7\linewidth]{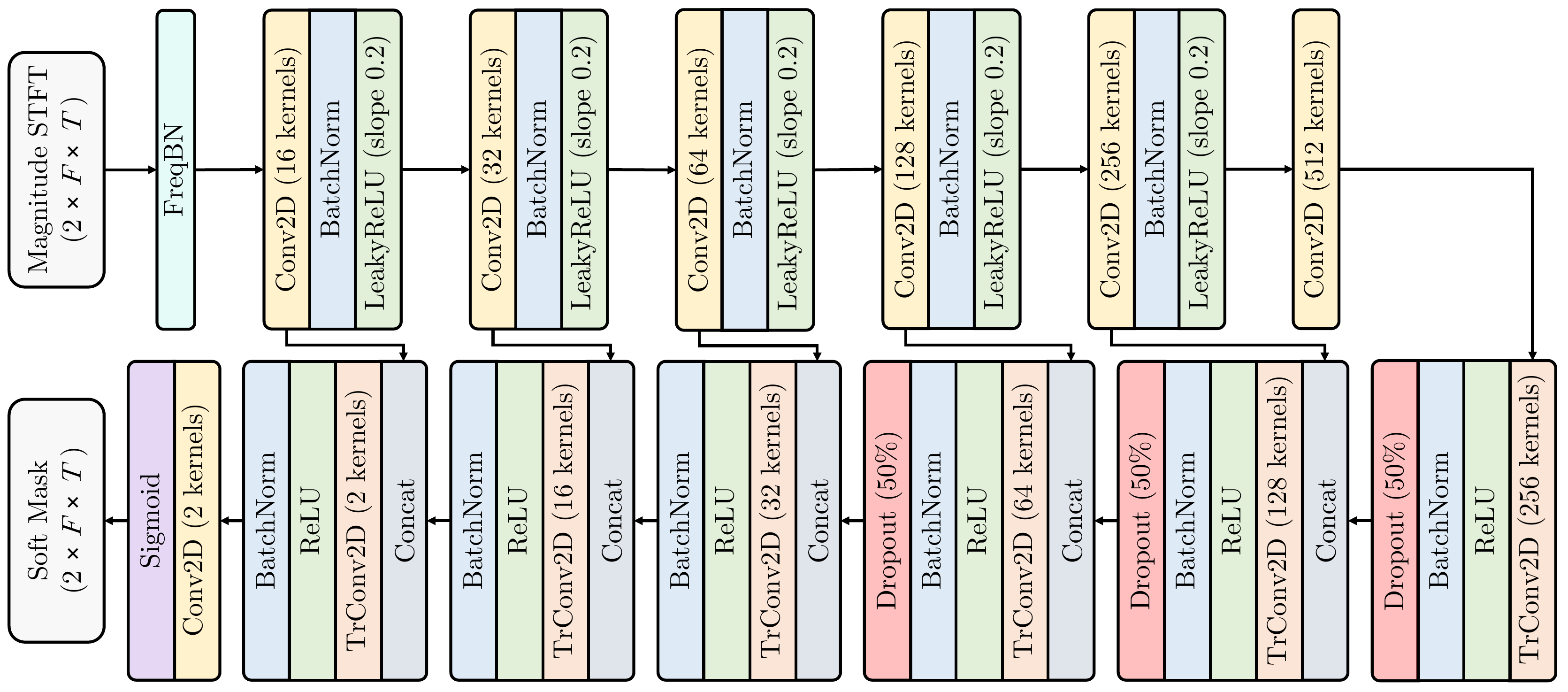}
 \caption{U-Net architecture.}
 \label{fig:unet}
\end{figure}

Inspired by prior work on MDX~\citep{Jansson2017SingingVS, hennequin2020spleeter, defossez2021hybrid}, LarsNet comprises five parallel U-Nets~\citep{ronneberger2015unet}, i.e., one for each target stem. 
As shown in Fig.~\ref{fig:larsnet}, the U-Nets operate in the time-frequency domain. Each is fed a (portion of a) two-channel magnitude STFT computed from the stereo waveform of a drum mixture and outputs a stem-specific soft mask of the same size. The time-domain stem signal is estimated by taking the inverse STFT (iSTFT) of the Hadamard product between the complex-valued STFT of the mixture and the real-valued mask thus obtained. The input magnitude STFT is first cropped to retain only the lowest $F$ frequency bands. Then, zero-padding is applied along the temporal dimension before segmenting it into an integer number of chunks of $T$ time frames. The resulting spectro-temporal patches $\mathbf{X}\in\mathbb{R}_{\ge 0}^{2\times F\times T}$ are normalized using Batch Norm computed across frequency bands (FreqBN), instead of across channels as it is usual in computer vision applications~\citep{ioffe2015batch}. 
As shown in Fig.~\ref{fig:unet}, each U-Net comprises $13$ convolutional layers. Such layers have $5\times 5$ kernels, $2\times 2$ stride, and $2\times 2$ padding, except for the last decoder layer which has $4\times 4$ kernels, $2\times 2$ dilation, and $3\times 3$ padding. The decoder yields a soft mask $\mathbf{M}\in[0,1]^{2\times F\times T}$  through a Sigmoid nonlinearity. The mask is zero-padded back to the original STFT size, and the temporal dimension is reinstated by concatenating all $T$-sized chunks.
Finally, the time-domain signal is reconstructed by taking the iSTFT of the masked magnitude after padding any frequency band above $F$ with zeros. Albeit more sophisticated phase estimation techniques exist, e.g., \citep{griffin_lim, kobayashi2022phase}, we opt to simply use the original mixture phase when transforming the signal back into the time domain for efficiency's sake.

\subsection{Model Training}
We train the U-Nets in parallel on five NVIDIA Titan V GPUs for $100\,000$ iterations using Adam, a learning rate of $0.0001$, and a batch size of $24$. Each mixture-stem training pair consists of aligned stereo waveform segments of $11.85$~s extracted from StemGMD with a stride of $2$~s, resulting in $110\,000$ segments per epoch. 
We compute the STFT using a periodic Hann window of length $4096$ and a hop size of $1024$ (ca.\ $23$~ms.) Since we are dealing with percussive instruments whose onsets have impulsive spectral characteristics, we set $F=2048$ to ensure a broadband estimate of every stem. Additionally, to avoid padding the STFT patches at training time, we fix $T=512$.
The loss function is computed as the $L^1$-norm of the error between the magnitude STFT patch of the ground truth stem $\mathbf{X}_i\in\mathbb{R}_{\ge 0}^{2\times F\times T}$ and that of the masked mixture, i.e.,
\begin{equation}
    \mathcal{L} = \lVert \mathbf{X}_i - \mathbf{M}_i \odot \mathbf{X} \rVert_1\,,
\end{equation}

\noindent where $\odot$ denotes the Hadamard product.
Throughout the training, we apply data augmentation as described in Section~\ref{sec:data_augmentation}.
We use six of the ten drum kits included in StemGMD to train LarsNet, whereas the remaining four are held out for evaluation purposes on unseen drum sounds (see Table~\ref{tab:sdr}.)

\subsection{Wiener Filtering}
Once the training is over, the magnitude STFT of each stem can be inferred independently of the others as 
\begin{equation}
    \hat{\mathbf{X}}_i = \mathbf{M}_i \odot \mathbf{X}\,.
\end{equation} 
Additionally, LarsNet implements $\alpha$-Wiener filtering~\citep{liutkus2015generalized} as in~\citep{dittmar2016amen}. Namely, the $i$th soft mask is computed as
\begin{equation}
    \tilde{\mathbf{M}}_i = \hat{\mathbf{X}}_i^\alpha \oslash \left(\sum_{j=1}^N \hat{\mathbf{X}}_j^\alpha + \epsilon\right)\,,
    \label{eq:alpha_wiener_filtering}
\end{equation}
and the magnitude STFT is estimated as $\tilde{\mathbf{X}}_i =\tilde{\mathbf{M}}_i \odot \mathbf{X}$,
where $\epsilon$ avoids division by zero, $\oslash$ indicates the Hadamard division, and the exponent $\alpha\in(0,2]$ is applied in an element-wise fashion. In the following, we report the results of this latter LarsNet variant by setting $\alpha=1$. Notice that, differently from the forward inference mechanism, \eqref{eq:alpha_wiener_filtering} combines the estimates of all $N=5$ stems, preventing full parallelization. 

\subsection{Target Stems}
In this work, we separate five stems (KD, SD, TT, HH, CY) even if StemGMD contains isolated tracks for nine different instruments (see Section \ref{sec:dataset}).
From an application standpoint, we argue that there are limited practical scenarios in which one might want to isolate, e.g., the High Tom from the Low-Mid Tom, as such a distinction mainly depends on the diameter and tuning of the drums that may vary greatly across drum kits. Furthermore, some drummers may use two tom-toms; others may have more than three. Having a single TT class allows one to isolate that entire family of drums regardless of the drum kit composition. Similarly, the CY class may be considered a stereo ``overhead'' track where all cymbals are recorded at once. In fact, the ride and crash cymbals are rarely recorded using dedicated microphones unlike, e.g., the hi-hat, which is typically close-miked. Finally, the hi-hat is a percussive instrument that can be played in two different ways (open or closed) depending on whether the pedal is pressed or not. In having a single HH class, we favored the consistency with respect to the acoustic source rather than the sounds it may produce, akin to having a microphone capturing the instrument.

\section{Data Augmentation}
\label{sec:data_augmentation}
Having access to isolated audio stems enables countless data augmentation strategies and allows one to draw inspiration from common music production practices. 
We propose to use the following six methods;\footnote{Other data augmentation methods were considered but not applied in the present study, such as \textbf{Panning (PN)}, i.e., randomly modify the level of the left and right channels of each stem according to a panning law, \textbf{Reverb (REV)}, i.e., stochastically apply artificial reverberation with random parameters to each stem independently of the others, and \textbf{Random (RND)}, i.e., select stems completely at random regardless of whether they come from the same drumming performance or not.} 
the order in which they are presented below corresponds to the order in which they are applied in our implementation.

\begin{enumerate}
\item \textbf{Kit-swap augmentation (KS)}: for each drum pattern, we create a novel mixture by adding together stems from randomly selected drum kits.

\item \textbf{Doubling augmentation (DB)}: in modern-day music production, layering multiple drum hits in a Digital Audio Workstation (DAW) is common practice. Inspired by this, for each stem, we compute a new track by averaging the same pattern from different drum kits. To save on I/O operations, this method is limited to two drum kits at a time.

\item \textbf{Pitch-shift augmentation (PS)}: to simulate a wider range of drums and cymbals, each stem is pitch shifted by a random amount of semitones using \texttt{SoX} as backend. Specifically, we randomly sample integer-valued shifts in the range of $\pm 3$ semitones.

\item \textbf{Saturation augmentation (ST)}: we apply nonlinear processing to individual stems to simulate compression and saturation, which are common when mixing drums in a DAW. Namely, for each stem, we compute the hyperbolic tangent ($\tanh$) after scaling the input waveform by a uniformly distributed random variable $\beta_i\sim\mathcal{U}_{[1,5]}$. 

\item \textbf{Channel-swap augmentation (CS)}: left and right channels of an audio track are randomly swapped.

\item \textbf{Remix augmentation (RX)}: Each stem is multiplied by a scalar $\gamma_i\sim\mathcal{U}_{\left[0.1, 1\right]}$,  corresponding to a gain variation in the range of $-20$~dB to $0$~dB. Albeit increasing the gain above $0$~dB would have been possible, we opted against it to avoid potential clipping. 

\end{enumerate}

\noindent If we apply all augmentations at once, the mixture signal 
\begin{equation}
    \mathbf{x}[n] = \sum_{i=1}^N \mathbf{x}_i[n]
\end{equation}
becomes

\begin{equation}
    \mathbf{x}_\text{aug}[n] = \sum_{i=1}^N \gamma_i \tanh\!\left(\beta_i\cdot \mathrm{shift}_{\pm 3}\!\left(\frac{\mathbf{C}\mathbf{x}_i^{(k)}[n] + \mathbf{C}\mathbf{x}_i^{(k')}[n]}{2}\right)\right)\!,
\end{equation}

\noindent where $i$ denotes the stem index out of $N$ stems, $k$ indicates the index of a randomly selected drum kit, $k\ne k'$, $\mathbf{C}=\begin{bsmallmatrix} 0 & 1\\ 1 & 0\end{bsmallmatrix}$, $\beta_i\sim\mathcal{U}_{\left[1, 5\right]}$, $\gamma_i\sim\mathcal{U}_{\left[0.1, 1\right]}$, and $\mathrm{shift}_{\pm 3}(\cdot)$ is an operator shifting the pitch of the input signal by a random integer amount in the range of $[-3, 3]$ semitones.
In practice, each augmentation method is applied stochastically and independently of the others. 
If applied, KS and RX involve all stems in the mixture, whereas applying DB, PS, ST, and CS depends on the outcome of a Bernoulli trial run for each voice. 
The probability of every augmentation is as follows: $\mathrm{Pr}(\mathrm{KS}) = \mathrm{Pr}(\mathrm{CS}) = 0.5$, and $\mathrm{Pr}(\mathrm{DB}) = \mathrm{Pr}(\mathrm{PS}) = \mathrm{Pr}(\mathrm{ST}) = \mathrm{Pr}(\mathrm{RX}) = 0.3$.
On top of that, at training time, we disable data augmentation
altogether
with a probability of 50\%.

\section{Evaluation}
To assess the performance of our deep DSS model, we consider the signal-to-distortion ratio ($\mathrm{SDR}$). In particular, we adopt the definition used in previous MDX challenges~\citep{mdx2021, mdx2023}, which~\citet{defossez2021hybrid} refers to as \nSDR.
Namely, for the $i$th stem, the metric is given by

\begin{equation}\label{eq:nsdr_i}
   \mathrm{nSDR}_i =  10 \log_{10} \dfrac{\sum_n \lVert\mathbf{x}_i[n]\rVert^2 + \epsilon}{\sum_n \lVert\mathbf{x}_i[n] - \hat{\mathbf{x}}_i[n]\rVert^2 + \epsilon}\,,
\end{equation}

\noindent where $n$ is the time index, $\epsilon=10^{-7}$ is a small scalar introduced to avoid numerical problems, $\mathbf{x}_i[n]$ is the $i$th ground truth source, and $\hat{\mathbf{x}}_i[n]$ is the estimated stereo waveform. Then, the overall \nSDR\ score is given by the average over 
the $N$ stems of every clip in the test set, namely 

\begin{equation}\label{eq:nsdr}
   \mathrm{nSDR} =  \frac{1}{N}\sum_{i=1}^N \mathrm{nSDR}_i\,.
\end{equation}

Different from the more widespread formulation by~\citet{vincent2006performance} (later referred to as $\mathrm{oSDR}$ for clarity's sake), \eqref{eq:nsdr} is simpler and faster to evaluate, which is convenient given the considerable size of large-scale deep learning datasets.
The $\mathrm{SDR}$ is well-defined as long as the discrete-time energy of the target signal $\mathbf{x}_i[n]$ is greater than zero. If not, the optimal ratio in \eqref{eq:nsdr_i}, i.e., assuming no distortion, is nonpositive and caps at $0$~dB. Hence, we divide the test set into two parts by annotating each stem according to the corresponding MIDI file. Specifically, we analyze the results for all those stems where at least one MIDI note is present (nonzero-energy stems) separately from those with no drum hits (zero-energy stems.)
We test LarsNet against the baseline methods described in Section~\ref{ssec:baseline}. The three methods are evaluated on the \textit{Eval Session} fold of StemGMD, comprising mixtures from ten drum kits, four of which were held out during training.

\subsection{Baseline Methods}
\label{ssec:baseline}
\citet{dittmar2014real} proposed a low-latency method for drums transcription and separation of KD, SD, and HH based on frame-wise nonnegative matrix factorization with semi-adaptive bases (SAB-NMF). 
The method aims at decomposing the magnitude STFT of the drum mixture into two nonnegative matrices, one comprising a set of spectral templates (bases) and the other containing the corresponding temporal activations. In particular, SAB-NMF achieves this factorization by processing each short-time spectrum independently of the others.

\citet{dittmar2016amen} proposed an alternative approach based on nonnegative matrix factor deconvolution (NMFD) followed by $\alpha$-Wiener filtering. Differently from other methods, the authors assume one has prior knowledge about the drum score and presents an \textit{informed} NMFD variant. However, such a prior is rarely available when it comes to real-life source separation problems. Hence, we implement the baseline decomposition 
from~\citep{dittmar2016amen},
which closely resembles the classic NMFD formulation by~\citet{smaragdis2004non}. 

Compared to~\citep{dittmar2014real} and \citep{dittmar2016amen}, we apply the multiplicative update rules~\citep{lee2000algorithms} for $K=200$ iterations instead of $25$ and~$30$, as this improved the \nSDR\ significantly.

In our implementation, the spectral basis functions are pre-computed using a partition of StemGMD containing isolated drum and cymbal hits, each synthesized at ten different velocities ranging from 30 to 127 using the same ten drum kits as in the main dataset.
Moreover, whereas the original references only considered mono files, we process the two stereo channels independently of one another. As for the remaining hyperparameters, we refer the reader to the original papers.

\subsection{Results}
Table~\ref{tab:sdr} reports the \nSDR\ for every drum kit and isolated stem in StemGMD \textit{Eval Session}. The ``All'' rows show the average \nSDR\ over all stems in a drum kit, while the ``All'' column lists the average across all drum kits for a given stem.

First, none of the methods exhibits a noticeable drop in performance when evaluated on the four held-out drum kits (marked with $^\dagger$ in Table~\ref{tab:sdr}), suggesting that all three methods can generalize in the face of unseen timbral characteristics.

By looking at the \nSDR\ of nonzero-energy stems, we notice that LarsNet provides, on average, a performance increment of $+20.39$~dB for KD, $+6.91$~dB for SD, $+20.63$~dB for TT, $+3.14$~dB for HH, and $+8.4$~dB for CY compared to SAB-NMF. Similarly, we report an increment of $+3.92$~dB for KD, $+11.84$~dB for SD, $+13.28$~dB for TT, $+3.33$~dB for HH, and $+9.78$~dB for CY with respect to NMFD. When accounting for all drum kits and stems, LarsNet yields an \nSDR\ of $17.7$~dB, against the $10.97$~dB of NMFD and $7.24$~dB of SAB-NMF. Respectively, the average $\mathrm{oSDR}$ scores~\citep{vincent2006performance} obtained using \texttt{mir\_eval}~\citep{raffel2014mir_eval} are $17.91$~dB for LarsNet, $11.02$~dB for NMFD, and $7.23$~dB for SAB-NMF, differing at most by $0.21$~dB from the respective \nSDR\ scores.

As for the results pertaining to zero-energy stems, both SAB-NMF and NMFD have an average \nSDR\ of $-25.56$~dB. 
This indicates a great deal of cross-talk between drum channels, i.e., sound components from other stems tend to leak into the TT, HH, and CY tracks. Conversely, LarsNet scores a perfect $0$~dB for many of the ten drum kits, meaning that the proposed method correctly outputs silence when no drum hits are present in the ground truth stem. Overall, LarsNet average \nSDR\ is $-0.84$~dB, corresponding to a remarkable $+24.7$~dB improvement upon the baselines.

\begin{table}[t]
    \centering
    \caption{nSDR computed on StemGMD Eval Session. 
    The six drum kits denoted by the superscript\;$^\star$\ were included in the training dataset, whereas the four drum kits denoted by\;$^\dagger$\ were held out for evaluation.}
    \label{tab:sdr}
    \resizebox{\linewidth}{!}{%
    \begin{tabular}{*{15}{c}}
            \toprule
            & \textbf{Method} & \textbf{Stem} & \textbf{No.} & \multicolumn{11}{c}{\textbf{Drum kit}} \\
            \midrule
             & & & & Brooklyn$^\star$ & East Bay$^\star$ & Heavy$^\star$ & Portland$^\star$ & Retro Rock$^\star$ & SoCal$^\star$ & Bluebird$^\dagger$ & Detroit Garage$^\dagger$ & Motown Rev.$^\dagger$ & Roots$^\dagger$ & All\\
            \midrule
            \multirow{21}{1.5em}{\begin{turn}{90}------------------ {Nonzero-energy stems} ------------------\end{turn}} & 
            \multirow{6}{*}{\textbf{SAB-NMF}~\citep{dittmar2014real}} & KD & $400$ ($100\%$) & $4.63$ & $3.48$ & $6.14$ & $9.57$ & $4.81$ & $5.49$ & $5.82$ & $5.30$ & $7.20$ & $15.55$ & $6.80$\\
            && SD & $400$ ($100\%$) & $13.03$ & $16.10$ & $17.56$ & $11.99$ & $18.32$ & $20.73$ & $16.86$ & $9.14$ & $10.27$ & $14.58$ & $14.86$\\
            && TT & $40$ ($10\%$) & $-11.66$ & $-11.81$ & $-12.13$ & $-10.36$ & $-14.78$ & $-10.06$ & $-11.45$ & $-16.96$ & $-11.06$ & $-5.07$ & $-11.53$\\
            && HH & $390$ ($97.5\%$) & $1.97$ & $6.50$ & $3.88$ & $-0.21$ & $6.61$ & $6.47$ & $1.14$ & $-3.34$ & $5.57$ & $4.26$ & $3.29$\\
            && CY & $50$ ($12.5\%$) & $-5.42$ & $-2.98$ & $-1.04$ & $-9.22$ & $-0.97$ & $-2.11$ & $-1.67$ & $-7.41$ & $-7.03$ & $-5.27$ & $-4.31$\\
            \cmidrule{3-15}
            && All & $1280$ & $5.54$ & $7.61$ & $8.17$ & $5.99$ & $8.74$ & $9.77$ & $7.01$ & $2.67$ & $6.54$ & $10.35$ & $7.24$\\
            \cmidrule{2-15}
            & \multirow{6}{*}{\textbf{NMFD}~\citep{dittmar2016amen}} 
            & KD & $400$ ($100\%$) & $21.98$ & $24.29$ & $12.33$ & $25.12$ & $24.67$ & $24.14$ & $28.26$ & $27.19$ & $23.08$ & $21.62$ & $23.27$\\
            && SD & $400$ ($100\%$) & $11.07$ & $10.83$ & $11.83$ & $10.27$ & $4.86$ & $12.75$ & $10.02$ & $8.75$ & $7.17$ & $11.71$ & $9.93$\\
            && TT & $40$ ($10\%$) & $-2.64$ & $-2.73$ & $-12.62$ & $-0.48$ & $-8.80$ & $-1.67$ & $0.23$ & $-1.85$ & $-5.55$ & $-5.70$ & $-4.18$\\
            && HH & $390$ ($97.5\%$) & $4.51$ & $3.35$ & $2.75$ & $-1.43$ & $4.21$ & $3.47$ & $3.72$ & $2.95$ & $3.93$ & $3.56$ & $3.10$\\
            && CY & $50$ ($12.5\%$) & $-6.49$ & $-5.34$ & $-5.02$ & $-7.37$ & $-3.81$ & $-4.08$ & $-6.96$ & $-7.16$ & $-3.40$ & $-7.30$ & $-5.69$\\
            \cmidrule{3-15}
            && All & $1280$ & $11.37$ & $11.70$ & $7.80$ & $10.32$ & $10.09$ & $12.38$ & $12.83$ & $11.79$ & $10.35$ & $11.04$ & $10.97$\\
            \cmidrule{2-15}
            & \multirow{6}{*}{\textbf{LarsNet}~(Ours)} 
            & KD & $400$ ($100\%$) & $27.07$ & $26.91$ & $26.22$ & $31.17$ & $27.10$ & $25.99$ & $29.29$ & $27.54$ & $25.56$ & $25.07$ & $\mathbf{27.19}$\\
            && SD & $400$ ($100\%$) & $21.48$ & $20.91$ & $20.84$ & $22.61$ & $23.07$ & $24.61$ & $23.01$ & $21.49$ & $17.42$ & $22.23$ & $\mathbf{21.77}$\\
            && TT & $40$ ($10\%$) & $9.22$ & $10.25$ & $9.49$ & $8.37$ & $10.43$ & $11.92$ & $9.50$ & $7.65$ & $8.37$ & $5.84$ & $\mathbf{9.10}$\\
            && HH & $390$ ($97.5\%$) & $6.80$ & $9.04$ & $5.69$ & $4.52$ & $7.70$ & $8.35$ & $6.03$ & $4.54$ & $5.59$ & $6.03$ & $\mathbf{6.43}$\\
            && CY & $50$ ($12.5\%$) & $4.46$ & $5.75$ & $5.08$ & $4.39$ & $4.82$ & $5.46$ & $3.91$ & $4.08$ & $3.32$ & $-0.35$ & $\mathbf{4.09}$\\
            \cmidrule{3-15}
            && All & $1280$ & $\mathbf{17.71}$ & $\mathbf{18.24}$ & $\mathbf{16.93}$ & $\mathbf{18.62}$ & $\mathbf{18.54}$ & $\mathbf{18.94}$ & $\mathbf{18.63}$ & $\mathbf{17.10}$ & $\mathbf{15.53}$ & $\mathbf{16.79}$ & $\mathbf{17.70}$\\
        \midrule
            \multirow{15}{1.5em}{\begin{turn}{90}--------- {Zero-energy stems} ---------\end{turn}} & 
            \multirow{4}{*}{\textbf{SAB-NMF}~\citep{dittmar2014real}} 
            & TT & $360$ ($90\%$) & $-37.33$ & $-37.09$ & $-36.77$ & $-36.38$ & $-33.89$ & $-30.49$ & $-35.29$ & $-33.49$ & $-32.85$ & $-26.06$ & $-33.96$\\
            && HH & $10$ ($2.5\%$) & $-27.91$ & $-30.16$ & $-28.13$ & $-20.68$ & $-20.11$ & $-25.25$ & $-25.13$ & $-27.85$ & $-24.16$ & $-27.55$ & $-25.69$\\
            && CY & $350$ ($87.5\%$) & $-20.60$ & $-14.13$ & $-7.98$ & $-25.97$ & $-12.57$ & $-11.15$ & $-10.92$ & $-23.23$ & $-25.06$ & $-17.44$ & $-16.91$\\
            \cmidrule{3-15}
            && All & $720$ & $-29.07$ & $-25.83$ & $-22.66$ & $-31.10$ & $-23.34$ & $-21.01$ & $-23.30$ & $-28.42$ & $-28.95$ & $-21.89$ & $-25.56$\\
            \cmidrule{2-15}
            & \multirow{4}{*}{\textbf{NMFD}~\citep{dittmar2016amen}} 
            & TT & $360$ ($90\%$) & $-27.19$ & $-24.14$ & $-33.91$ & $-20.97$ & $-35.82$ & $-23.25$ & $-24.46$ & $-23.34$ & $-27.14$ & $-30.62$ & $-27.08$\\
            && HH & $10$ ($2.5\%$) & $-24.67$ & $-28.37$ & $-27.26$ & $-25.54$ & $-22.82$ & $-23.34$ & $-21.59$ & $-23.88$ & $-26.43$ & $-26.67$ & $-25.06$\\
            && CY & $350$ ($87.5\%$) & $-24.61$ & $-23.53$ & $-21.61$ & $-28.85$ & $-21.13$ & $-22.20$ & $-24.15$ & $-24.44$ & $-25.51$ & $-24.12$ & $-24.01$\\
            \cmidrule{3-15}
            && All & $720$ & $-25.90$ & $-23.90$ & $-27.84$ & $-24.87$ & $-28.50$ & $-22.74$ & $-24.27$ & $-23.88$ & $-26.33$ & $-27.41$ & $-25.56$\\
            \cmidrule{2-15}
            & \multirow{4}{*}{\textbf{LarsNet}~(Ours)} 
            & TT & $360$ ($90\%$) & $0.00$ & $0.00$ & $0.00$ & $0.00$ & $0.00$ & $0.00$ & $0.00$ & $0.00$ & $0.00$ & $-8.99$ & $\mathbf{-0.90}$\\
            & & HH & $10$ ($2.5\%$) & $0.00$ & $-4.34$ & $-3.78$ & $0.00$ & $0.00$ & $0.00$ & $0.00$ & $-7.24$ & $-5.48$ & $-23.65$ & $\mathbf{-4.45}$\\
            & & CY & $350$ ($87.5\%$) & $0.00$ & $0.00$ & $0.00$ & $0.00$ & $0.00$ & $0.00$ & $0.00$ & $0.00$ & $-2.46$ & $-4.33$ & $\mathbf{-0.68}$\\
            \cmidrule{3-15}
            & & All & $720$ & $\mathbf{0.00}$ & $\mathbf{-0.06}$ & $\mathbf{-0.05}$ & $\mathbf{0.00}$ & $\mathbf{0.00}$ & $\mathbf{0.00}$ & $\mathbf{0.00}$ & $\mathbf{-0.10}$ & $\mathbf{-1.27}$ & $\mathbf{-6.93}$ & $\mathbf{-0.84}$\\
        \bottomrule
        \end{tabular}%
    }%
\end{table}

\subsection{Real-Time Performance}
LarsNet is implemented in Python using PyTorch. 
Despite not being optimized for speed, the model achieves an average Real-Time Ratio (RTR) of 
$0.016$ on a single NVIDIA TITAN V GPU ($62.5$ times faster than real-time) and $0.15$ on an Intel Xeon E5-2687W ($6.6$ times faster than real-time). Conversely, on the same CPU, SAB-NMF and NMFD achieve an RTR of $4.85$ (slower than real-time) and $0.4$ (only $2.5$ times faster than real-time), respectively.

\section{Conclusions}
In this manuscript, we presented StemGMD, the first large-scale audio dataset of isolated drum stems encompassing all the percussion instruments of a canonical nine-piece acoustic drum kit, and LarsNet, a first-ever deep drum source separation model. Based on a parallel arrangement of dedicated U-Nets, LarsNet can extract five stems from a stereo drum mixture up to $62.5$ times faster than real-time. At the same time, LarsNet is shown to significantly outperform state-of-the-art methods based on nonnegative spectro-temporal decomposition in terms of signal-to-distortion ratio and unwanted cross-talk artifacts. 
Ultimately, LarsNet is intended to serve as a baseline for future work on drum source separation, while StemGMD has the capability to foster further research into the applications of deep learning to music demixing.

\bibliographystyle{unsrtnat}
\bibliography{article}  






\end{document}